\documentclass[twocolumn,amsmath,amssymb,showpacs,letterpaper,superscriptaddress]{revtex4-2}

\usepackage{graphicx,xspace,subfigure}
\usepackage{dcolumn}
\usepackage{bm}
\usepackage[dvipsnames]{xcolor}
\usepackage{hyperref}
\hypersetup{
   colorlinks=true,
   linkcolor=blue,
   filecolor=magenta,      
   urlcolor=blue,
   citecolor=blue,
}
\usepackage{float}
\usepackage{amsmath}
\usepackage{gensymb}
\usepackage{chngcntr}
\usepackage[toc,page]{appendix}

\usepackage{chemformula}
\usepackage{siunitx}

\usepackage{pstricks}
\usepackage{dutchcal}
\usepackage{psfrag}

\usepackage[normalem]{ulem}
\usepackage{appendix}

\usepackage{upgreek}

\renewcommand{\bm}[1]{\boldsymbol{\mathbf{#1}}}
\newcommand{\ud}{\mathrm{d}}
\newcommand{\bra}{\left\langle}
\newcommand{\ket}{\right\rangle}
\newcommand{\re}{\operatorname{Re}}
\newcommand{\im}{\operatorname{Im}}
\newcommand{\pSHG}{p_{\text{SHG}}}
\newcommand{\np}{\mathcal{p}}
\newcommand{\ie}{i.e.}
\newcommand{\GammaSHG}{\Gamma_{\text{SHG}}}

\newcounter{tempa}
\newcounter{tempb}
\newcounter{tempc}
\newcounter{tempd}

\newenvironment{ddiag}[1]{\psset{unit=1.3mm,fillstyle=solid,fillcolor=white}
   \begin{pspicture}[shift=-5](0,-7)(#1,7)}{\end{pspicture}
}

\newcommand{\ggmoy}[3]{\psline[linewidth=0.5](#1,#3)(#2,#3)}
\newcommand{\eemoy}[3]{\psline[linewidth=0.5,linestyle=dashed](#1,#3)(#2,#3)}

\newcommand{\gggmoy}[4]{\psline[linewidth=0.5](#1,#3)(#2,#4)}

\newcommand{\pparticule}[2]{\pscircle(#1,#2){1}}
\newcommand{\nnonlineaire}[2]{
   \setcounter{tempa}{#1}
   \addtocounter{tempa}{-1}
   \setcounter{tempb}{#2}
   \addtocounter{tempb}{-1}
   \setcounter{tempc}{#1}
   \addtocounter{tempc}{1}
   \setcounter{tempd}{#2}
   \addtocounter{tempd}{1}
   \psframe(\value{tempa},\value{tempb})(\value{tempc},\value{tempd})
}

\newcommand{\ccorreldeuxc}[4]{
   \psline[linestyle=dashed](#1,#2)(#3,#4)
}

\newenvironment{ddddiag}[1]{\psset{unit=1.3mm,fillstyle=solid,fillcolor=white}
   \begin{pspicture}[shift=-8](0,-10)(#1,10)}{\end{pspicture}
}

\begin{document}

\title{Photon diffusion in space and time in a second-order nonlinear disordered medium}

\author{Rabisankar Samanta}
\affiliation{Nano-optics and Mesoscopic Optics Laboratory, Tata Institute of Fundamental Research,
             1 Homi Bhabha Road, Mumbai, 400 005, India}
\author{Romain Pierrat}
\email{romain.pierrat@espci.psl.eu}
\affiliation{Institut Langevin, ESPCI Paris, PSL University, CNRS, 1 rue Jussieu,
             75005 Paris, France}
\author{Rémi Carminati}
\affiliation{Institut Langevin, ESPCI Paris, PSL University, CNRS, 1 rue Jussieu,
             75005 Paris, France}
\affiliation{Institut d’Optique Graduate School, Université Paris-Saclay, F-91127 Palaiseau, France}
\author{Sushil Mujumdar}
\email{mujumdar@tifr.res.in}
\affiliation{Nano-optics and Mesoscopic Optics Laboratory, Tata Institute of Fundamental Research,
             1 Homi Bhabha Road, Mumbai, 400 005, India}
\date{\today}

\begin{abstract}
   We report experimental and theoretical investigations on photon diffusion in a second-order nonlinear disordered
   medium under conditions of strong nonlinearity. Experimentally, photons at the fundamental wavelength
   ($\lambda=\SI{1064}{nm}$) are launched into the structure in the form of a cylindrical pellet, and the
   second-harmonic ($\lambda=\SI{532}{nm}$) photons are temporally analyzed in transmission.  For comparison, separate
   experiments are carried out with incident green light at $\lambda=\SI{532}{nm}$. We observe that the second harmonic
   light peaks earlier compared to the incident green photons.  Next, the sideways spatial scattering of the fundamental
   as well as second-harmonic photons is recorded.  The spatial diffusion profiles of second-harmonic photons are seen
   to peak deeper inside the medium in comparison to both the fundamental and incident green photons. In order to give
   more physical insights into the experimental results, a theoretical model is derived from first principles.  It is
   based on the coupling of transport equations. Solved numerically using a Monte Carlo algorithm and experimentally
   estimated transport parameters at both wavelengths, it gives excellent semi-quantitative agreement with the
   experiments for both fundamental and second-harmonic light. 
\end{abstract}

\maketitle

\section{Introduction}\label{intro}

Electromagnetic wave propagation and scattering are ubiquitous in diverse fields such as optics, condensed matter
physics, biology, atmospheric optics, etc~\cite{ishimaru1978}. Among the various related phenomena, diffusion of light
has attracted maximum attention in the last decades~\cite{CARMINATI-2021-1}. Two factors have primarily motivated these
studies, namely, the occurrence of diffusion of light in tissues~\cite{martelli2009}, and the parallels of light
diffusion with electron propagation in disordered conductors~\cite{kittel2012}. The first sub-field, namely, diffusion
in tissues, has led to significant signs of progress in imaging of inclusions in living media~\cite{tuchin2007}, with an
aim to replace hazardous high-energy radiation. The second domain overlaps with condensed matter studies and is intended
to understand mesoscopic effects in light transport in disordered media.  Indeed, studies on mesoscopic optics have led
to better insights on electronic transport, such as the transition from the diffusion regime to localization
regime~\cite{Anderson1958} upon a sufficient increase in disorder.

A primary reason for the success of these studies in the optical domain has been the inherent noninteracting nature of
the photons, as well as the possibility to directly image the intensity distribution, which makes the analyses easier
than for electrons. Interestingly, photon propagation allows for additional aspects of transport, such as amplification
or nonlinearity. On the one hand, coherent amplification accompanied by diffusion or localization has led to the
creation of novel optical sources named random lasers~\cite{letokhov1968, lawandy1994, pradhan1994, wiersma2008,
tiwari2013, uppu2015}. On the other hand, studies coupling nonlinearity with the disorder have attracted growing
attention~\cite{melnikov2004, boer1993, qiao2019, frohlich1986, mckenna1992,
schwartz2007,fishman2012,lahini2008,heiderich1995, wellens2005,agranovich1988, yoo1989, simon1992, wang1993}. For
instance, a large effort has been focused on the effects of $\chi^{(3)}$ nonlinearity on Anderson
localization~\cite{frohlich1986, mckenna1992, schwartz2007}, wherein the question of whether nonlinearity subjugates
localization is still an open question~\cite{fishman2012}. Experimental attempts to address this problem in
one-dimensional systems showed that the localized mode continues to exist under introduced
nonlinearity~\cite{lahini2008}. Theoretically, the effects of $\chi^{(3)}$ nonlinearity on coherent backscattering (or
weak localization) have also been documented~\cite{heiderich1995, wellens2005}. Likewise,  $\chi^{(2)}$ materials, in
particular relevance to second harmonic generation (SHG)~\cite{franken1961}, have been studied in the regimes of
diffusion and weak localization~\cite{agranovich1988, yoo1989, simon1992, wang1993}. Specifically, several consequences
of random quasi-phase-matching in disordered $\chi^{(2)}$ materials have been discussed in recent years~\cite{savo2020,
muller2021, morandi2022}. Furthermore, speckle dynamics in these materials have been recently quantified and
theoretically modeled~\cite{samanta2020, samanta2022}. Interestingly, second harmonic generation in a diffusive material
creates a unique paradoxical situation of incoherent transport, i.e. diffusion, coupled with an inherently coherent
phenomenon of second harmonic generation, and hence is of fundamental research interest. Over the years, a few
theoretical~\cite{kravtsov1991, makeev2003}  and experimental reports~\cite{faez2009} have addressed this peculiar
scenario. In one of the earliest reports, Kravtsov \textit{et al.}~\cite{kravtsov1991} theoretically studied two models
of disorder, namely point scatterers in a nonlinear medium and grainy nonlinear scatterers, and observed sharp peaks in
the angular distribution of backward diffuse second harmonic light. In another approach, Makeev \textit{et al.}
theoretically studied the diffusion of second-harmonic (SH) light in a colloidal suspension of spherical nonlinear
particles,  wherein they found that the average SH intensity was independent of the linear scattering properties of the
medium~\cite{makeev2003}. In the experimental effort on GaP powders, a consistent picture that described the
second-harmonic intensity distribution in the sample was obtained via the diffusion equation, which invoked nonlinearity
as a conversion rate~\cite{faez2009}.

The various models described above successfully invoked the scenario of weak nonlinearity, wherein the nonlinearity did
not affect the distribution of fundamental photons. However, these works focused on the spatial behavior of fundamental
and second-harmonic light, while no attention was paid to the temporal behavior, under ultrashort pulse illumination. In
this work, we address this question, and others, in the light of experiments and of a theoretical model. We carry out
spatial and temporal investigations of light propagating through a strongly disordered KDP (potassium dihydrogen
phosphate) pellet. We diagnose pulse propagation through the system and measure the transport mean free paths for the
fundamental and second harmonic light using the transmitted pulses. Subsequently, we measure the longitudinal spatial
distribution of both components. To give more physical insights into the processes at play, we build from first
principles a theoretical model involving the coupling of two Radiative Transfer Equations (RTE). Solved numerically
using a Monte Carlo scheme, the model gives reliable results compared to the experiment. In Sec.~\ref{smaples_experiments}, we
introduce the samples and the experiments and we present all experimental results. In Sec.~\ref{theory}, we present the
theoretical model as well as the associated numerical simulations in the same geometry as in the experiments. The
discussion and conclusion are given in Sec.~\ref{conclusion}.

\section{Samples and experiments}\label{smaples_experiments}

\begin{figure}[htbp]
   \centering
   \includegraphics[width=\linewidth]{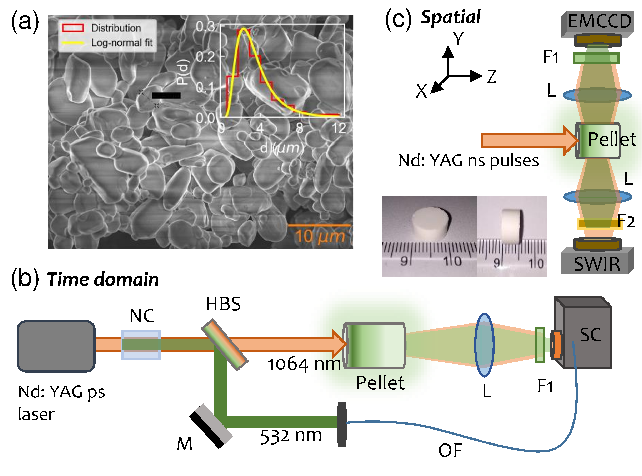}
   \caption{(a) SEM image of particles. The scale bar is \SI{10}{\micro m}. Inset: Distribution of particle size. (b) Schematic
   for temporal measurements. Notations: NC, Nonlinear Crystal; HBS, Harmonic Beam Splitter; M, Mirror; L, Lens; F1,
   Filter (for $\lambda=\SI{532}{nm}$); OF, Optical Fiber; SC, Streak Camera. (c) Schematic of the experiment on spatial
   diffusion. Notations: L, Lens; F1; F2, Filters; SWIR, Short-wavelength Infra-Red; EMCCD, Electron Multiplying
   Charge Coupled Device. Inset images show the pellet with \SI{1}{cm} diameter and length of \SI{5}{mm}.}
   \label{fig:setup}
\end{figure}

We exploited commercially available KDP (@ EMSURE ACS) as our nonlinear material. It has a large second-order optical
nonlinearity ($\sim\SI{0.43}{pm/V}$). A fine powder of KDP was realized after $10$ minutes of ball milling.
Figure~\ref{fig:setup}(a) depicts the scanning electron micrograph image of KDP grains while the inset shows the size
distribution of the KDP grains, with grain sizes in the range of \SIrange{2}{8}{\micro m}. The size distribution (bar
graph, red) is approximately log-normal (yellow line), peaking at \SI{3.11}{\micro m}. A pellet of length $L=\SI{5}{mm}$
and radius $a=\SI{0.5}{cm}$ was created under a hydraulic press. The pellet was baked to remove remnant moisture at a
temperature of \SI{80}{\degree C}, which was far below the transition temperature (\SI{190}{\degree C}, \cite{itoh1975})
of KDP. Before going to the main experiment, we evaluated the effective refractive indices and internal reflection
coefficients to be used later as input parameters in our theoretical model.  Using the Maxwell-Garnett formula
\cite{choy1999} (see App.~\ref{int_RC} regarding the use of the Marxell-Garnett formula), the real parts of the
effective refractive indices of the pellet were found to be $n_r(2\omega)=1.45$ and $n_r(\omega)=1.43$ at
$\lambda=\SI{532}{nm}$ and $\lambda=\SI{1064}{nm}$ respectively. The internal power reflection coefficients at
$\lambda=\SI{532}{nm}$ and $\lambda=\SI{1064}{nm}$ are estimated to be $R(2\omega)=0.53$ and $R(\omega)=0.52$
respectively~\cite{zhu1991} (see App.~\ref{int_RC} for the detailed derivation of the internal reflection coefficient). 

A schematic of the experimental setups is depicted in Fig.~\ref{fig:setup}\,(b) and (c). Two different experiments were
performed. For the temporal measurements [Fig.~\ref{fig:setup}\,(b)], Nd:YAG ps laser pulses (EKSPLA, PL2143B), with
pulse width $\sim\SI{30}{ps}$ at the fundamental wavelength of $\lambda=\SI{1064}{nm}$ (hereafter referred to as IR) and
beam waist $\sim\SI{5}{mm}$, were made incident on the front face of the pellet. The second harmonic generated light
(hereafter referred to as SHG) that was transmitted from the back face was directed into the Streak Camera (Optronics
SC-10). The streak camera possesses a temporal resolution of less than 3 ps and its specified spectral sensitivity
ranges from 200 to \SI{950}{nm}. The spectral range of the Streak Camera did not allow for measuring the IR pulse. A small
part of the second harmonic light generated by a clear nonlinear crystal was directed toward the marker pulse input of
the Streak Camera to calibrate the zero of the time axis. For the spatial measurements [Fig.  \ref{fig:setup}(c)], the
pellet was pumped with the fundamental of the Nd: YAG ns laser (pulse width $\sim\SI{5}{ns}$, repetition rate
\SI{10}{Hz}). The pulsed IR beam was launched normally onto the front face of the pellet. The scattered IR photons were
imaged from one side of the pellet by a combination of lens and InGaAs CCD, as shown in Fig. \ref{fig:setup}.
Simultaneously, a Silicon CCD along with a lens was employed to measure the internally generated and scattered second
harmonic photons (SHG). For comparison to be used later, in both the temporal and spatial experiments, we also
externally launched frequency-doubled green light from the source laser ($\lambda=\SI{532}{nm}$, hereafter referred to
as incident green or IG), whose temporal transmitted profile and spatial scattering profile were measured separately by
the same Streak Camera and Silicon CCD respectively. The CCD images provided the spatial variation of light intensity
along the length of the pellet.

Our sample consists of a large number of KDP microcrystals with random orientations. The incident photons experience
multiple scattering due to the refractive index mismatch between the microcrystals and the background medium. SHG
photons are generated at random positions within the medium by the crossing of two IR photons and also undergo multiple
scattering. At large optical thickness and long times, we can expect the transmitted multiply scattered IR and SHG
intensities to follow a diffusion process, as confirmed by the spatial intensity profiles shown below.

\begin{figure}[h]
   \centering
   \includegraphics[width=\linewidth]{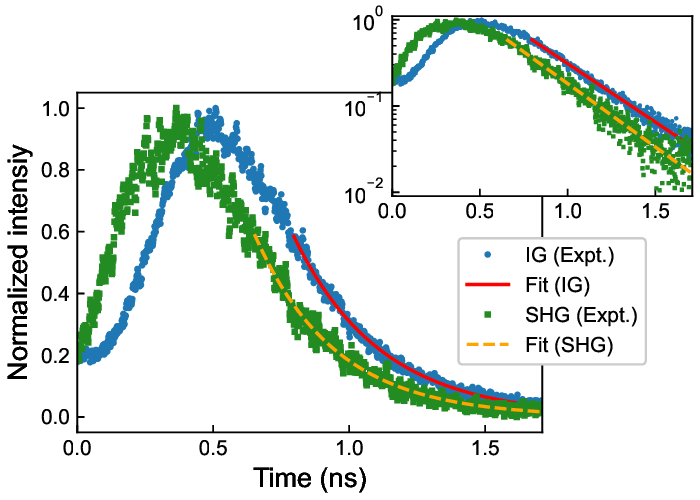}
   \caption{Experimental temporal diffusion profiles of SHG (green squares) and IG (blue dots) in a pellet with
   $L=\SI{5}{mm}$. The SHG signal peaks earlier in time. The orange dashed and red solid lines fit the exponential
   decays for SHG and IG respectively. Inset: diffusion profiles on the semilog scale show comparable decay lengths.}
   \label{fig:temporal}
\end{figure}

The experimental temporal profiles of SHG (green squares) and IG (blue dots) in the pellet, plotted in
Fig.~\ref{fig:temporal}, exhibit a classic diffusive behavior. We observe an interesting trend in the transmission where
SHG photons have shorter residence times than IG photons, given that the SHG profile peaks earlier. The temporal
profiles of SHG and IG show clear exponentially decaying tails, as evidenced in the inset on a logarithmic scale.
Interestingly, the decay rates for the two curves are also very similar, a fact which we will return to in our
theoretical work. The tails were fit by the expression $A\exp(-\Gamma t)$, with $\Gamma$ as the decay rate. The
estimated decay rates fitted to numerous temporal profiles of the same sample amount to $3.17 \pm 0.13\,\text{ns}^{-1}$
and $3.25 \pm 0.11\,\text{ns}^{-1}$ for the SHG and IG, respectively. These values are as close as a fitting routine can
provide. The apparent deviation may therefore be probably due to the strong noise level. The orange dashed and red solid
lines in Fig. \ref{fig:temporal} correspond to the fits for SHG and IG respectively. Nonetheless, we emphasize here
that, only the IG data can be fit \emph{a priori} using diffusion theory. Although we use the same analysis for the SHG
light, the fact that it follows a similar diffusion equation cannot be justified based on an existing theory. The theory
developed in the next section will elaborate on this part and also justify the comparable decays seen in the
experiments. In the presence of absorption and side loss, the decay rate $\Gamma$ in a cylindrical system can be written
as~\cite{gaikwad2014}
\begin{equation}\label{exp_decay_rate}
    \Gamma(2\omega)=\mathcal{D}(2\omega)\left[\frac{\alpha_1^2}{a_e(2\omega)^2}+\frac{\pi^2}{L_e(2\omega)^2}+\frac{3}{\ell_t(2\omega)\ell_a(2\omega)}\right]
\end{equation}
where the diffusion coefficient is given by
\begin{align}
\mathcal{D}(2\omega)=\frac{v_E(2\omega)\ell_t(2\omega)}{3}.
\end{align}
For non-resonant scatterers, the energy velocity can be well approximated by $v_E(2\omega)=c/n_r(2\omega)$ where $c$ is
the light velocity in vacuum. $\ell_t$ and $\ell_a$ are the transport and absorption mean free paths respectively.
$a_e$ and $L_e$ are the effective radius and thickness of the pellet respectively and are given by
$a_e(2\omega)=a+z_e(2\omega)$ and $L_e(2\omega)=L+2z_e(2\omega)$ where $z_e$ is the extrapolation length given by
\begin{align}
   z_e(2\omega)=\frac{2}{3}\ell_t(2\omega)\left[\frac{1+R(2\omega)}{1-R(2\omega)}\right].
\end{align}
$\alpha_1$ in Eq.\ref{exp_decay_rate} is the first zero of the zero-order Bessel function $\operatorname{J}_0$.  By
fitting the experimentally measured decay rates at $\lambda=\SI{532}{nm}$ for different sample lengths with
Eq.~(\ref{exp_decay_rate}), we extract $\ell_t(2\omega)\approx\SI{32}{\micro m}$ and $\ell_a(2\omega)\approx\SI{11}{mm}$
(side loss). Due to the limitation of spectral sensitivity, the streak camera could not be used for the temporal profile
of the IR light and the determination of the transport mean free path. We derived the value of $\ell_{t}(\omega)$ from a
coherent backscattering experiment~\cite{maret1987}. The estimated value of $\ell_{t}(\omega)$ is $\sim \SI{216}{\micro
m}$. Note that for both the harmonics, $\{L, a\}\gg\ell_t$ and $t\gg v_E\ell_t$. This signifies that our sample is
clearly in the diffusive regime~\cite{zhang1999}.

\begin{figure}[h]
   \centering
   \includegraphics[width=0.8\linewidth]{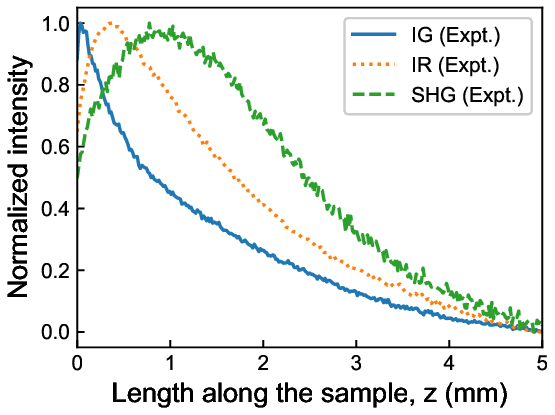}
   \caption{Experimental spatial intensity profiles of incident green (IG, blue solid line), fundamental infrared (IR, orange
   dotted line), and second-harmonic generated green (SHG, green dashed line) light for the $L=\SI{5}{mm}$ pellet. The SHG signal peaks much
   deeper in the sample, and IG peaks closest to the entry face inside the pellet, followed by IR.}
   \label{fig:spatial}
\end{figure}

Next, we examine the spatial diffusion profiles in the pellet. Figure~\ref{fig:spatial} displays the experimentally
measured spatial profiles for the IG (blue solid line), IR (orange dotted line), and SHG (green dashed line) photons.
Each profile shows a characteristic diffusive peak inside the input edge. The first peak to appear is the IG one
followed by the IR one. This behavior is expected since $\ell_t(2\omega)<\ell_t(\omega)$. The SHG peak may be expected
to appear at a location where the fundamental intensity is maximum. However, this is not the case, and the maximum
intensity of the SHG is observed deeper in the sample where the IR is about \SI{80}{\%} of its peak intensity. This
comes from the fact that the SHG beam is generated in the medium by the IR beam and then propagates with the same
transport characteristics as the IG beam. This behavior will be well reproduced by the theoretical model. Regarding the
wings deep inside the sample, an exponential decay is observed. This arises from the loss mechanisms at play in the
sample coming from the finite transverse extent of the pellet, the linear absorption, and the nonlinearity in the case
of the IR light. 

\section{Theoretical model}\label{theory}

\subsection{Disorder model}\label{disorder}

In order to give more physical insights, we develop a theoretical model taking into account multiple scattering and
second harmonic generation. To this end, we generalize the standard multiple scattering theory to derive transport
equations for the fundamental and second harmonic intensities averaged over the configurations of the disorder. This
derivation is similar to that of Ref.~\onlinecite{samanta2022}. However, there are three major differences: (1) we
explicitly consider the time-dependent regime, (2) we drop the intensity decorrelation due to scatterer displacements
and (3) we take absorption into account. This is important to note that this does not change the way the derivation is
performed since these differences do not enter directly the computation of the second harmonic phase function. The
interested reader can refer to App.~\ref{app_theory} where the derivation is given. We present in the following only the
assumptions and the main results.  We first define a model of the disorder. The real material is made of a fine KDP
powder containing particles of different shapes and sizes.  Therefore, the most simple and natural disorder model
consists of a continuous and complex fluctuating permittivity $\epsilon(\bm{r})$. The disorder microstructure is then
characterized by a spatial correlation function chosen to be Gaussian, in the form
\begin{equation}\label{disorder_correlation}
   C_{\epsilon}(\bm{r}-\bm{r}',\omega)=\bra\delta\epsilon(\bm{r},\omega)\delta\epsilon^*(\bm{r}',\omega)\ket
      =|\Delta\epsilon(\omega)|^2C(\bm{r}-\bm{r}')
\end{equation}
with
\begin{equation}
   C(\bm{r}-\bm{r}')=\exp\left[-\frac{|\bm{r}-\bm{r}'|^2}{2\ell^2}\right].
\end{equation}
In the equations above, $\bra\ldots\ket$ represents an average over all disorder configurations (statistical average).
$\delta\epsilon(\bm{r},\omega)=\epsilon(\bm{r},\omega)-\bra\epsilon(\bm{r},\omega)\ket$ is the
fluctuating part of the permittivity, $|\Delta\epsilon(\omega)|^2$ is the amplitude of the correlation and $\ell$ is the
correlation length. $|\Delta\epsilon(\omega)|^2$ depends on frequency since the permittivity $\epsilon$ is
dispersive. However, $\ell$ involves only the geometrical structure of the disorder and thus does not depend on
frequency. This also implies that the $\chi^{(2)}$ nonlinearity is supposed to be correlated in a similar way (\ie, with
the same correlation length). We thus have
\begin{equation}\label{chi_correlation}
   C_{\chi}(\bm{r}-\bm{r}',\omega)=\bra\chi^{(2)}(\bm{r},\omega)\chi^{(2)*}(\bm{r}',\omega)\ket
      =|\Delta\chi(\omega)|^2C(\bm{r}-\bm{r}').
\end{equation}

\subsection{Transport equations}

We first consider the case of the fundamental beam at frequency $\omega$ corresponding to $\lambda=\SI{1064}{nm}$ and also
denoted by IR. The application of the standard multiple scattering theory leads to the no less standard Radiative
Transfer Equation (RTE) given by~\cite{CHANDRASEKHAR-1950}
\begin{multline}\label{linear_rte}
   \left[\frac{1}{v_E(\omega)}\frac{\partial}{\partial t}
      +\bm{u}\cdot\bm{\nabla}_{\bm{r}}+\frac{1}{\ell_e(\omega)}\right]I(\bm{r},\bm{u},t,\omega)
\\
      =\frac{1}{\ell_s(\omega)}\int p(\bm{u}\cdot\bm{u}',\omega)I(\bm{r},\bm{u}',t,\omega)\ud\bm{u}'.
\end{multline}
Although the pellet is large enough, a diffusion-type equation is not adequate enough to describe the experimental
results. Indeed, this kind of equation does not allow an accurate reconstruction of the fluxes at short times and/or at
small depths. A transport type equation such as Eq.~(\ref{linear_rte}) is required.
In this equation, $I(\bm{r},\bm{u},t,\omega)$ is the specific intensity, that can be seen as the radiative flux at
position $\bm{r}$, in direction $\bm{u}$, at time $t$ and at frequency $\omega$. More precisely, the derivation
from the first principles
shows that it is given by the Wigner transform of the field. It reads
\begin{multline}\label{specific_intensity}
   \delta(k-k_r)I(\bm{r},\bm{u},t,\omega)=\int \bra E\left(\bm{r}+\frac{\bm{s}}{2},\omega+\frac{\Omega}{2}\right)\right.
\\
\times
     \left. E^*\left(\bm{r}-\frac{\bm{s}}{2},\omega-\frac{\Omega}{2}\right)\ket
       e^{-ik\bm{u}\cdot\bm{s}-i\Omega t}\ud\bm{s}\frac{\ud\omega}{2\pi}
\end{multline}
where $E$ is the electric field, $E^*$ being its complex conjugate counterpart. In this definition, the electric field
is a scalar quantity. Indeed, we choose here to neglect the effect of the polarization which is a very good
approximation in the multiple scattering regimes where the field can be considered to be fully
depolarized~\cite{VYNCK-2014}. $k_r(\omega)=n_r(\omega)k_0$ is the real part of the effective wavevector where
$k_0=\omega/c$ is the wavenumber in vacuum. We recall that $n_r$ is the real part of the effective refractive index. It
describes the phase velocity of the average field $\bra E\ket$ (\ie, the field averaged over all possible configurations
of the disorder). $\ell_e(\omega)$ and $\ell_s(\omega)$ are the extinction and scattering mean-free paths respectively.
They are connected by the relation $\ell_e(\omega)^{-1}=\ell_s(\omega)^{-1}+\ell_a(\omega)^{-1}$ with $\ell_a(\omega)$
the absorption mean free path.  $p(\bm{u},\bm{u}',\omega)$ is the phase function representing the fraction of power
scattered from direction $\bm{u}'$ to direction $\bm{u}$ during a single scattering event. For the Gaussian disorder
considered here, it is given by
\begin{multline}\label{linear_phase_function}
   p(\bm{u},\bm{u}',\omega)\propto\np(|k_r(\omega)\bm{u}-k_r(\omega)\bm{u}'|)
\\
   \text{where}\quad
   \np(q)=\exp\left[-\frac{q^2\ell^2}{2}\right]
\end{multline}
and normalized such that $\int p(\bm{u},\bm{u}',\omega)\ud\bm{u}'=1$. As previously mentioned, $v_E(\omega)$ is the
energy velocity, well approximated by the phase velocity for non-resonant scatterers. Equation~(\ref{linear_rte}) can be
easily interpreted using a random walk picture. In this picture, light undergoes a random walk whose average step is given by the
scattering mean free path $\ell_s(\omega)$ and whose angular distribution at each scattering event is given by the phase
function $p(\bm{u},\bm{u}',\omega)$. The absorption is taken into account along a path of length $s$ through an
exponential decay $\exp[-s/\ell_a(\omega)]$.

Let us focus now on the second harmonic light at frequency $2\omega$ corresponding to $\lambda=\SI{532}{nm}$ and also
denoted by SHG. By generalizing the standard multiple scattering theory taking into
account second harmonic generation, we obtain a RTE given by (see App.~\ref{app_theory} for details)
\begin{multline}\label{non_linear_rte}
   \left[\frac{1}{v_E(2\omega)}\frac{\partial}{\partial t}
      +\bm{u}\cdot\bm{\nabla}_{\bm{r}}+\frac{1}{\ell_e(2\omega)}\right]I(\bm{r},\bm{u},t,2\omega)
\\
      =\frac{1}{\ell_s(2\omega)}\int p(\bm{u}\cdot\bm{u}',2\omega)I(\bm{r},\bm{u}',t,2\omega)\ud\bm{u}'
\\
      +\alpha\iint \pSHG(\bm{u},\bm{u}',\bm{u}'',\omega)I(\bm{r},\bm{u}',t,\omega)
\\\times
      I(\bm{r},\bm{u}'',t,\omega)\ud\bm{u}'\ud\bm{u}''.
\end{multline}
This equation is very similar to Eq.~(\ref{linear_rte}) except for the presence of a non-linear source term.
Because of the statistical average over disorder configurations, it can be shown that the dominant contribution in
terms of paths for the second harmonic beam involves two SHG processes (one to generate the field at $2\omega$ and the
second to generate its complex conjugate) at the same location and time~\cite{samanta2022}. This implies that there is
no phase shift to take into account between two fields, one at frequency $\omega$ and the second at frequency $2\omega$
and there is no phase-matching condition as in standard SHG experiments without scattering. Thus the SHG process reduces
to the product of two specific intensities at
$\omega$ coming from two different directions $\bm{u}'$ and $\bm{u}''$. $\alpha$ is a factor containing all constants
involved in the SHG process such as $\chi^{(2)}$. $\pSHG(\bm{u},\bm{u}',\bm{u}'',\omega)$ is the SHG phase function
describing the distribution of the SHG source term in direction $\bm{u}$ arising from two linear specific intensities
coming from directions $\bm{u}'$ and $\bm{u}''$. In the case of the correlated disorder we consider here, it is given by
\begin{equation}
   \pSHG(\bm{u},\bm{u}',\bm{u}'',\omega)\propto\np(|k_r(2\omega)\bm{u}-k_r(\omega)\bm{u}'-k_r(\omega)\bm{u}''|)
\end{equation}
where the $\np$ function is given in Eq.~(\ref{linear_phase_function}), and normalized such that $\int
\pSHG(\bm{u},\bm{u}',\bm{u}'',\omega)\ud\bm{u}'\bm{u}''=1$. Thus, the SHG phase function is directly related to the
disorder correlation function in the same way as the standard phase function since the SHG and scattering processes both
take place in the scatterers. We note that these equations are obtained in the perturbative approach meaning that the
beam at $\omega$ is a source term for the beam at $2\omega$ but the beam at $2\omega$ has a negligible impact on the
beam at $\omega$.

\subsection{Numerical simulations}

\begin{figure}[!htb]
   \centering
   \psfrag{a}[c]{(a)}
   \psfrag{b}[c]{(b)}
   \psfrag{n}[c]{$n=1$}
   \psfrag{m}[c]{$n_r$}
   \psfrag{L}[c]{$L$}
   \psfrag{r}[c]{$a$}
   \psfrag{z}[c]{$z$}
   \includegraphics[width=\linewidth]{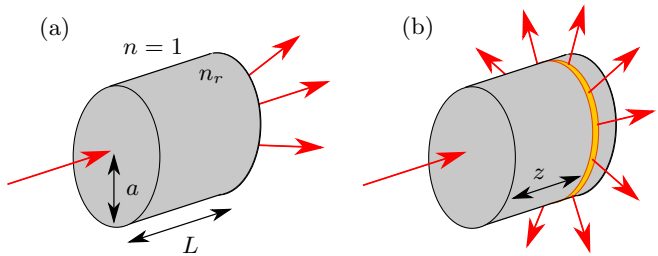}
   \caption{Geometry considered in the Monte Carlo simulations. The illumination is a Gaussian beam with waist $w_i$
   impacting the front face of the pellet under normal incidence. The detection is performed through the back face of
   the pellet [(a), temporal simulation] or along the side face [(b), spatial simulation]. The cylinder contains the
   powder (pellet) and has a radius $a$ and a height $L$. The refractive index mismatch between the cylinder (real part 
   of the effective refractive index $n_r$) and the external medium assumed to be air is also taken into account in
   the simulation.}
   \label{fig:numerical_geometry}
\end{figure}

Equations~(\ref{linear_rte}) and (\ref{non_linear_rte}) are solved using a Monte Carlo scheme in geometries as close as
possible to that in the real experiments (see Fig.~\ref{fig:numerical_geometry} for details).  In particular, the
crystal grains have sizes ranging from $\SI{2}{\micro m}$ to $\SI{8}{\micro m}$, which is large compared to the
wavelength. The correlation length is taken such that $k_0\ell=3$ for all simulations. Using this value and the
experimental estimate of the real part $n_r$ of the effective refractive index, we get estimates for the anisotropy
factors $g(\omega)=0.95$ and $g(2\omega)=0.99$. Then the values of the scattering mean-free paths are deduced from the
experimental estimates of the transport mean-free paths using the relation $\ell_s=\ell_t(1-g)$. Experimental estimates
are directly used for the absorption mean-free paths $\ell_a$. This shows that the disorder correlation length $\ell$
impacts only $\ell_s$. Recalling that $\ell_t$ are small compared to the dimensions of the pellet, the diffusive regime
applies. Thus $\ell_t$ is much more important than $\ell_s$ at large depth and time and the real value of $k_0\ell$
weakly impacts the numerical results except at small depth and time where slight variations can be visible. In practice,
three simulations are performed. The first is done to compute a 6D map (3 positional, 2 directional, and 1 temporal
coordinates) of the specific intensity at $\omega$. It gives also access to the linear detected flux at $\omega$ denoted
by $F_{\text{IR}}(t)$. The second is done to evaluate the SHG detected flux at $2\omega$ [\ie, $F_{\text{SHG}}(t)$]
using the previous map as a source term. And the last is performed to estimate the linear detected flux at $2\omega$
[\ie, $F_{\text{IG}}(t)$].

\begin{figure}[ht]
   \centering
   \includegraphics[width=\linewidth]{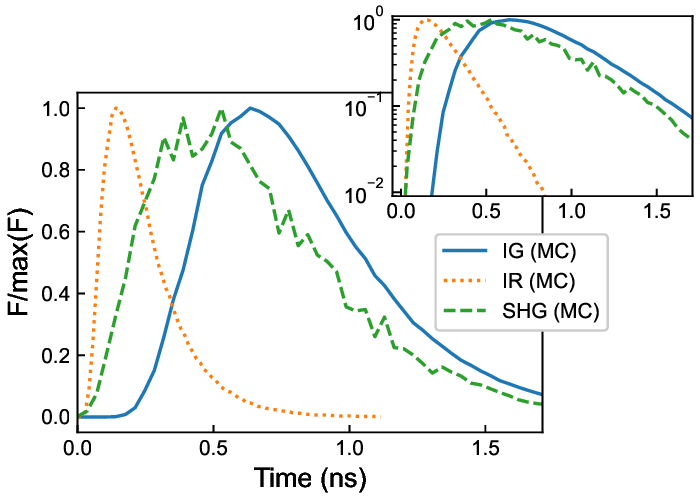}
   \caption{Numerically simulated temporal transmitted flux profiles of incident green (IG, blue solid line),
   fundamental infrared (IR, orange dotted line), and second-harmonic generated green (SHG, green dashed line) light for
   the $L=\SI{5}{mm}$ pellet.  Inset: flux profiles on a semilog scale.  We also observe a poorer SNR on the SHG curve.
   This is due to slow numerical convergence (only a fraction of the SHG light is collected on the detector, since the
   source term for the SHG beam is dispersed throughout the pellet and a large part of the SHG beam emerges from the
   other sides of the pellet). All parameters are chosen to mimic the experiment. The pellet is illuminated from the
   front face with an ultrashort Gaussian pulse of normalized waist $k_0w_i=\num{2.95e4}$ (at frequency $\omega$) and
   $k_0w_i=\num{1.77e4}$ (at frequency $2\omega$) and under normal incidence. The detection is performed through the
   back face of the pellet [see Fig.~\ref{fig:numerical_geometry}\,(a)]. The other parameters are given by
   $k_0a=\num{2.96e4}$, $k_0L=\num{2.96e4}$, $k_0\ell_s(\omega)=\num{1.28e3}$, $k_0\ell_s(2\omega)=\num{1.89e2}$,
   $k_0\ell_a(\omega)=\num{1e10}$, $k_0\ell_a(2\omega)=\num{5.9e5}$, $n_r(\omega)=1.43$, $n_r(2\omega)=1.45$ and
   $k_0\ell=3$.}
   \label{fig:numerical_results_time}
\end{figure}

\begin{figure}[ht]
   \centering
   \includegraphics[width=0.8\linewidth]{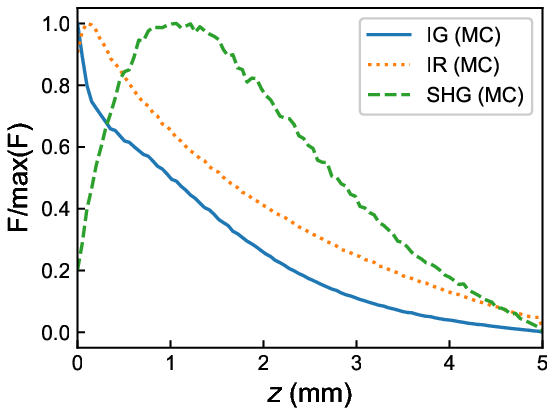}
   \caption{Numerically simulated spatial flux profiles of incident green (IG, blue solid line), fundamental infrared (IR,
   orange dotted line), and second-harmonic generated green (SHG, green dashed line) light for the $L=\SI{5}{mm}$ pellet. Numerical
   diffusion profiles show excellent agreement with the experimental results in Fig.~\ref{fig:spatial}.  All
   parameters are chosen to mimic the experiment. The pellet is illuminated from the front face with a Gaussian beam of
   normalized waist $k_0w_i=\num{2.95e4}$ (at frequency $\omega$) and $k_0w_i=\num{1.77e4}$ (at a frequency $2\omega$)
   and under normal incidence. The detection is performed along the side face of the pellet at different normalized
   depths $k_0d$ [see Fig.~\ref{fig:numerical_geometry}\,(b)]. The other parameters are given by  $k_0a=\num{2.96e4}$,
   $k_0L=\num{2.96e4}$, $k_0\ell_s(\omega)=\num{1.28e3}$, $k_0\ell_s(2\omega)=\num{1.89e2}$,
   $k_0\ell_a(\omega)=\num{1e10}$, $k_0\ell_a(2\omega)=\num{5.9e5}$, $n_r(\omega)=1.43$, $n_r(2\omega)=1.45$ and
   $k_0\ell=3$.}
   \label{fig:numerical_results_space}
\end{figure}

The results are shown in Figs.~\ref{fig:numerical_results_time} and \ref{fig:numerical_results_space} for the temporal
and spatial profiles respectively. The specific parameters for each experiment are detailed in the captions and are
chosen to mimic the experiments. Both figures show very good agreement with the experiment. Let us first
consider the temporal results of Fig.~\ref{fig:numerical_results_time}. We clearly recover the correct exponential time
decay of the flux at long times which is a feature of the diffusive regime. In particular, the long time-dependence of
the flux at the fundamental frequency $F_{\text{IR}}(t)$ is given by $\exp(-\Gamma(\omega)t)$. In the diffusive regime,
since all radiative quantities have the same time dependence, the specific intensity itself is also decaying according
to the same law. This implies that the source term of the non-linear RTE has a long time dependence given by
$\exp(-2\Gamma(\omega)t)$. By convolving this source term with the time response of a short pulse, we simply deduce the
long time behavior of the SHG flux at $2\omega$ which is given by
\begin{align}\nonumber
   F_{\text{SHG}}(t) & \propto \int_0^t e^{-2\Gamma(\omega)t'}e^{-\Gamma(2\omega)(t-t')}\ud t'
\\\label{time_decay}
         & \propto \begin{cases}
                     \displaystyle\frac{e^{-\Gamma(2\omega)t}-e^{-2\Gamma(\omega)t}}{2\Gamma(\omega)-\Gamma(2\omega)}
                        & \text{if $\Gamma(2\omega)\ne 2\Gamma(\omega)$},
                   \\
                     \displaystyle te^{-\Gamma(2\omega)t} & \text{otherwise}.
                   \end{cases}
\end{align}
For the powder considered here, $2\Gamma(\omega)>\Gamma(2\omega)$, which means that the long time behavior of the SHG
flux is given by $e^{-\Gamma(2\omega)t}$. In other words, the IG and SHG beams have the same time decay at long times.
This is clearly seen in Fig.~\ref{fig:numerical_results_time}. However, we note that a different behavior could have
been observed depending on the order relation between $2\Gamma(\omega)$ and $\Gamma(2\omega)$. We also note that
Eq.~(\ref{time_decay}) has been obtained under the assumption of exponential decay at all times for all quantities
included in this calculation. Since this is valid only at long times, slight deviations could be observed numerically
when comparing the IG and SHG beams. However, Fig.~\ref{fig:numerical_results_time} shows that this is not the case. A
quick comparison with Figure \ref{fig:temporal} reveals very good qualitative agreement with the experiments.

Regarding the spatial distribution of intensity displayed in Fig.~\ref{fig:numerical_results_space}, we clearly see that
the green light in the linear domain shows a maximum right at the input interface of the sample. The fundamental IR
light, similarly, shows a maximum slightly deeper in the sample, obviously owing to the larger value of $\ell_t$, which
is the consequence of the larger wavelength. The generated second-harmonic light peaks much deeper into the sample,
reflecting the process of local generation and subsequent diffusion of the emitted light. These trends accurately
reproduce the experimental observations in Fig.~\ref{fig:spatial} in a quantitative manner, certifying the completeness
of the numerical model. Overall, for the total spatio-temporal behavior, we can claim the agreement to be
semi-quantitative.

\section{Summary and Conclusions}\label{conclusion}

In conclusion, we have experimentally and numerically investigated the diffusion of light in a second-order nonlinear
disordered material with a strong nonlinear coefficient. In the experiments, a cylindrical pellet of KDP microcrystals,
packed in a random orientation, was employed. The spatial and temporal diffusion profiles of fundamental and second
harmonic light have been measured, along with incident light at the second harmonic wavelength.  The experimental data
were used to estimate the transport parameters to the best possible accuracy. Next, the experimental results were
supported by a theoretical model derived from first principles and led to the coupling of transport equations for the
linear and non-linear beams. A Monte Carlo scheme has been used to solve this model numerically in the same geometry as
in the experiment. Excellent agreement has been obtained in the spatial behavior seen in the experiments, along with
very good qualitative agreement in the temporal behavior. Moreover, this model allowed a clear interpretation of the
experimental results. 

We note that this work addresses spatial and temporal diffusion separately for both fundamental and second-harmonic
light. In the domain of linear disorder, researchers have investigated spatio-temporal diffusion \cite{pattelli2016,
sperling2016}, which has very interesting implications in light transport. Naturally, spatio-temporal diffusion of
second-harmonic light, simultaneously with the fundamental, will form an exciting frontier in nonlinear disorder
physics. Indeed, the streak camera used for this work has the capability to simultaneously capture spatial and temporal
diffusion. Besides, in our cylindrical geometry, there is a scope for exploration of transverse diffusion of second
harmonic as well as fundamental light. These topics will form the object of our future interest, starting with the
current numerical models. We expect the present work to lay down the platform for further modeling of light transport in
nonlinear disorder.

\section*{Acknowledgments}

R.S. and S.M. acknowledge the support from Kartik K. Iyer and Sanjay K. Upadhyay during the sample-making. We also thank
Bhagyashri Chalke for her help in the SEM measurements. R.S. appreciates the support from Ashwin K. Boddeti for the
initial experiments. R.S. thanks Sandip Mondal for the fruitful discussions. We express our gratitude towards  the
Department of Atomic Energy, Government of  India for funding for Project Identification No. RTI4002 under the DAE OM
No 1303/1/2020/R\&D-II/DAE/5567 dated 20.8.2020; Swarnajayanti Fellowship, Department of Science and Technology,
Ministry of Science and Technology, India. This work has also received support under the program “Investissements
d’Avenir” launched by the French Government. The authors declare no conflicts of interest.

\appendix

\section{Determination of the effective refractive indices and internal reflection coefficients}\label{int_RC}

\subsection{\texorpdfstring{Determination of the real part of the effective refractive index $n_r$ of the random medium}{}}

In the pellet, the volume fraction of the KDP is $0.88$.
The Maxwell Garnett equation reads~\cite{choy1999}
\begin{equation}
    \left( \frac{\varepsilon_r-\varepsilon_m}{\varepsilon_r+2\varepsilon_m} \right)
      =\delta_i \left( \frac{\varepsilon_i-\varepsilon_m}{\varepsilon_i+2\varepsilon_m}\right)
\end{equation}
where $ \varepsilon_r$ is the real part of the effective dielectric constant of the medium, $\varepsilon_i$ of the inclusions,
and $\varepsilon_m$ of the matrix; $\delta_i$ is the volume fraction of the inclusions. In our case, the air percentage
is smaller than the KDP powder. We can therefore assume that air is the inclusion and KDP is the matrix. Putting
$\delta_i=0.12$, $\varepsilon_i=1$ and $\varepsilon_m=n_{m}^2=(1.5124)^2$ we get $n_r(2\omega)=1.45$ at
$\lambda=\SI{532}{nm}$. At $\lambda=\SI{1064}{nm}$, the refractive index of KDP is $1.4938$~\cite{kdp}. The calculated value of
$n_r(\omega)$ is $1.4336$.

\subsection{\texorpdfstring{Determination of the internal reflection coefficient $R$}{}}

We can obtain an estimate of the reflection coefficient using Fresnel's law. We assume that the direction and
polarization of diffusing light incident on the boundary from inside the sample are completely random and the sample
surface is flat. For an angle of incidence $\theta$, the power reflection coefficient $R(\theta)$ averaged over
polarization is
\begin{equation}
   R(\theta)=\frac{R_{\perp}(\theta)+R_{\parallel}(\theta)}{2}
\end{equation}
where
\begin{align}
   R_{\perp}(\theta)=\left|\frac{n_1\cos\theta-n_2\sqrt{1-\left(\frac{n_1}{n_2}\sin\theta\right)^2}}
      {n_1\cos\theta+n_2\sqrt{1-\left(\frac{n_1}{n_2}\sin\theta\right)^2}}\right|^2
\end{align}
and
\begin{align}
   R_{\parallel}(\theta)=\left|\frac{n_1\sqrt{1-\left(\frac{n_1}{n_2}\sin\theta\right)^2}-n_2\cos\theta}
      {n_1\sqrt{1-\left(\frac{n_1}{n_2}\sin\theta\right)^2}+n_2\cos\theta}\right|^2
\end{align}
are the Fresnel power reflection coefficients for incident light polarized perpendicular and parallel to the plane
of incidence, respectively. Then the power internal reflection coefficient $R$ averaged
over the incident angle $\theta$ is given by~\cite{zhu1991} 
\begin{equation}
   R=\frac{3C_2+2C_1}{3C_2-2C_1+2}
\end{equation}
where
\begin{align}
   C_1=\int_{0}^{\frac{\pi}{2}} R(\theta) \sin\theta \cos\theta \ud\theta
\end{align}
and
\begin{align}
   C_2=\int_{0}^{\frac{\pi}{2}} R(\theta) \sin\theta \cos^2\theta \ud\theta.
\end{align}
Putting $n_1=n_r$ and $n_2=1$, we get $R(\omega)=0.52$ (\ie{} for $\lambda=\SI{1064}{nm}$) and $R(2\omega)=0.53$
(\ie{} for $\lambda=\SI{532}{nm}$).

\section{Derivations of the transport equations}\label{app_theory}

This appendix is dedicated to the derivations of the transport equations at $\omega$ and $2\omega$. We report here the
main steps and we detail the computation of the SHG source term for the non-linear transport equation at $2\omega$. The
derivations rely mainly on the standard multiple scattering theory, details of which can be found in various
books and reviews~\cite{RYTOV-1989,APRESYAN-1996,ROSSUM-1999,CARMINATI-2021-1}.

\subsection{Scattering potential and Green function}

One of the main building blocks of the multiple scattering theory is the scattering potential defined by
\begin{equation}
   V(\bm{r},\omega)=k_0^2\left[\epsilon(\bm{r},\omega)-\epsilon_b(\omega)\right]
\end{equation}
where $\epsilon_b(\omega)$ is the permittivity of the background homogeneous medium (reference medium). In order to
ensure that the perturbative method applied afterward is as accurate as possible, it is advisable to choose for
$\epsilon_b$ the statistical average of the permittivity (\ie, $\bra\epsilon(\bm{r},\omega)\ket$) which leads to
\begin{equation}
   V(\bm{r},\omega)=k_0^2\delta\epsilon(\bm{r},\omega).
\end{equation}
This potential describes the scattering process by the heterogeneities of the medium.
The second ingredient is the background Green function that connects two consecutive scattering events. It is given by
\begin{equation}\label{green_function}
   G_b(\bm{r}-\bm{r}',\omega)=\frac{\exp(ik_b|\bm{r}-\bm{r}'|)}{4\pi|\bm{r}-\bm{r}'|}
\end{equation}
where $k_b=n_bk_0$ is the wavevector in the background medium. We note that this Green function is a scalar quantity
here since we choose to neglect the polarization. Indeed, it can be shown that the polarization is completely washed out
on average after a propagation distance of the order of the transport mean-free path inside a disordered
medium~\cite{VYNCK-2014}. In particular, this expression given by Eq.~(\ref{green_function}) corresponds to the Green
function of the scalar homogeneous wave equation given by
\begin{equation}
   \Delta G_b(\bm{r}-\bm{r}',\omega)+k_b^2G_b(\bm{r}-\bm{r}',\omega)=-\delta(\bm{r}-\bm{r}').
\end{equation}

\subsection{Self-energy and intensity vertex}

Two operators are used to describe the propagation of the field in disordered media: the self-energy
$\Sigma(\bm{r},\bm{r}',\omega)$ entering the Dyson equation that governs the average field propagation and the intensity
vertex $\Gamma(\bm{r},\bm{r}',\bm{\uprho},\bm{\uprho}',\omega)$ entering the Bethe-Salpeter equation that governs the
field-field correlation evolution. Both contain an infinite series of scattering sequences that are statistically not
factorizable. Considering a dilute medium (quantified further by the condition $k_r\ell_e\gg 1$), these scattering sequences can
be seen as Taylor expansions. Thus, we can apply a perturbative approach and truncate
the series to the first non-zero order. In the following, we consider only the frequency $\omega$ but the same result holds at $2\omega$ (for the linear beams only).

For the self-energy $\Sigma$, this gives
\begin{equation}
   \Sigma(\bm{r},\bm{r}')=\bra V(\bm{r},\omega)G_b(\bm{r}-\bm{r}',\omega)V(\bm{r}',\omega)\ket_c
\end{equation}
where $\bra\cdot\ket_c$ represents a statistical average restricted to the connected part, \ie $\bra VG_bV\ket_c=\bra
VG_bV\ket - \bra V\ket G_b\bra V\ket$. Since the disorder correlation function $C$ depends only on
$|\bm{r}-\bm{r}'|$ (statistical homogeneity and isotropy), we have in the Fourier domain
\begin{equation}
   \Sigma(\bm{k},\bm{k}',\omega)=8\pi^3\delta(\bm{k}-\bm{k}')\widetilde{\Sigma}(k,\omega).
\end{equation}
This allows us to define the real part of the effective optical index by
\begin{equation}
   n_r(\omega)=n_b+\frac{\re\widetilde{\Sigma}(k_r,\omega)}{2n_rk_0^2}
\end{equation}
where we recall that $k_r(\omega)=n_r(\omega)k_0$ and $k_0=\omega/c$. This is a closed equation and the computation of
$n_r$ in practice is not an easy task except in very dilute media. That being known, the extinction mean-free path is defined by
\begin{equation}\label{le}
   \frac{1}{\ell_e(\omega)}=\frac{\im\widetilde{\Sigma}(k_r,\omega)}{k_r}.
\end{equation}

For the intensity vertex $\Gamma$, we have
\begin{equation}\label{gamma}
   \Gamma(\bm{r},\bm{r}',\bm{\uprho},\bm{\uprho}',\omega)
      =\bra V(\bm{r},\omega) V^*(\bm{\uprho},\omega)\ket_c \delta(\bm{r}'-\bm{r}')\delta(\bm{\uprho}-\bm{\uprho'}).
\end{equation}
By invoking one again the statistical homogeneity and isotropy, we get into the Fourier domain
\begin{equation}
   \widetilde{\Gamma}(\bm{k},\bm{k}',\bm{\upkappa},\bm{\upkappa}',\omega)
      =8\pi^3\delta(\bm{k}-\bm{k}'-\bm{\upkappa}+\bm{\upkappa}')
         \Gamma(\bm{k},\bm{k}',\bm{\upkappa},\bm{\upkappa}',\omega)
\end{equation}
which leads to the following expressions of the phase function and of the scattering mean-free path:
\begin{equation}
   \frac{1}{\ell_s(\omega)}p(\bm{u},\bm{u}',\omega)
      =\frac{1}{16\pi^2}\widetilde{\Gamma}(k_r\bm{u},k_0\bm{u}',k_0\bm{u},k_0\bm{u}',\omega)
\end{equation}
with the normalization
\begin{equation}
   \int p(\bm{u},\bm{u}',\omega)\ud\bm{u}'=1.
\end{equation}
This finally leads to
\begin{align}\label{ls}
   \frac{1}{\ell_s(\omega)}
      & =\frac{k_0^2|\Delta\epsilon|^2}{16\pi^2}\int_{4\pi}C(\bm{q})\ud\bm{u}'
\\
   p(\bm{u},\bm{u}',\omega)
      & =C(\bm{q})\left[\int_{4\pi}C(\bm{q})\ud\bm{u}'\right]^{-1}
\\\label{phase_function}
   & = \frac{k_r^2\ell^2\exp[-q^2\ell^2/2]}{2\pi\left[1-\exp(-2k_r^2\ell^2)\right]}
\end{align}
with $\bm{q}=k_r(\bm{u}-\bm{u}')$ and $C(\bm{q})$ the Fourier transform of the disroder correlation function $C$. In the
case of a non-absorbing medium where the imaginary part of the permittivity vanishes, we have from Eqs.~(\ref{le})
and (\ref{ls})
\begin{equation}\label{la}
   \frac{1}{\ell_e(\omega)}=\frac{1}{\ell_s(\omega)}
      =\frac{k_0^2|\Delta\epsilon|^2}{16\pi^2}\int_{4\pi}C(k_r(\bm{u}-\bm{u}'))\ud\bm{u}'
\end{equation}
which implies that $\ell_a^{-1}(\omega)=0$. In practice for the numerical simulations performed in this study, the
parameters $n_r$, $\ell_e$ and $\ell_s$ are not computed using $\Sigma$ and $\Gamma$ but are given by experimental
measurements. However, Eq.~(\ref{phase_function}) is used for the profile of the phase function.

\subsection{Transport equation in the linear regime}

Let us first consider the transport equation at $\omega$ in the linear regime. Considering a dilute medium such that
$k_r\ell_e(\omega)\gg 1$, we can show that the dominant term in the field-field correlation function is given by the
well-known ladder diagram:
\begin{equation}\label{ladder}
   \begin{ddiag}{32}
      \rput(0,3){$E(\bm{r},\omega)$}
      \rput(0,-3){$E^*(\bm{r},\omega)$}
      \ggmoy{5}{10}{-3}
      \ggmoy{5}{10}{3}
      \ggmoy{10}{16}{-3}
      \ggmoy{10}{16}{3}
      \ggmoy{16}{22}{-3}
      \ggmoy{16}{22}{3}
      \eemoy{22}{27}{-3}
      \eemoy{22}{27}{3}
      \ccorreldeuxc{10}{3}{10}{-3}
      \ccorreldeuxc{16}{3}{16}{-3}
      \ccorreldeuxc{22}{3}{22}{-3}
      \pparticule{10}{-3}
      \pparticule{10}{3}
      \pparticule{16}{-3}
      \pparticule{16}{3}
      \pparticule{22}{-3}
      \pparticule{22}{3}
      \rput(30,3){$E_0$}
      \rput(30,-3){$E_0^*$}
   \end{ddiag}.
\end{equation}
In this representation, the top line represents a path for the electric field $E$ and the bottom line is for a path of
its complex conjugate $E^*$. Solid and dashed thick lines correspond to average Green functions (describing propagation
between consecutive scattering events) and average fields respectively. Circles denote scattering events and vertical
dashed lines represent statistical correlations between scattering events through Eq.~(\ref{gamma}). This diagram
shows that the field-field correlation function evolution essentially reduces to the propagation of an intensity. This
implies that the field-field correlation function is governed by a transport equation, namely the Radiative Transfer
Equation (RTE) that reads
\begin{multline}
   \left[\frac{1}{v_E(\omega)}\frac{\partial}{\partial t}
      +\bm{u}\cdot\bm{\nabla}_{\bm{r}}+\frac{1}{\ell_e(\omega)}\right]I(\bm{r},\bm{u},t,\omega)
\\
      =\frac{1}{\ell_s(\omega)}\int p(\bm{u},\bm{u}',\omega)I(\bm{r},\bm{u}',t,\omega)\ud\bm{u}'
\end{multline}
which is Eq.(\ref{linear_rte}).

\subsection{Transport equation in the second harmonic regime}

Since we consider a perturbative approach for the SHG, the main point to address in that case is the expression of the
source term. We can show theoretically and validate numerically~\cite{samanta2022} that the leading diagram is given by
\begin{equation}\label{non_linear_ladder}
   \begin{ddddiag}{60}
      \rput(0,3){$E(\bm{r},2\omega)$}
      \rput(0,-3){$E^*(\bm{r},2\omega)$}
      \ggmoy{5}{11}{-3}
      \ggmoy{11}{17}{-3}
      \ggmoy{17}{23}{-3}
      \ggmoy{23}{29}{-3}
      \ggmoy{29}{35}{-3}
      \ggmoy{35}{41}{-3}
      \ggmoy{41}{47}{-3}
      \eemoy{47}{53}{-3}
      \ggmoy{5}{11}{3}
      \ggmoy{11}{17}{3}
      \ggmoy{17}{29}{3}
      \ggmoy{29}{35}{3}
      \ggmoy{35}{41}{3}
      \ggmoy{41}{47}{3}
      \eemoy{47}{53}{3}
      \ccorreldeuxc{11}{-3}{11}{3}
      \ccorreldeuxc{17}{-3}{17}{3}
      \ccorreldeuxc{23}{-3}{23}{3}
      \ccorreldeuxc{29}{-3}{29}{3}
      \ccorreldeuxc{35}{-3}{35}{3}
      \ccorreldeuxc{41}{-3}{41}{3}
      \ccorreldeuxc{47}{-3}{47}{3}
      \pparticule{11}{-3}
      \pparticule{17}{-3}
      \pparticule{23}{-3}
      \pparticule{35}{-3}
      \pparticule{41}{-3}
      \pparticule{47}{-3}
      \pparticule{11}{3}
      \pparticule{17}{3}
      \pparticule{23}{3}
      \pparticule{35}{3}
      \pparticule{41}{3}
      \pparticule{47}{3}
      \gggmoy{29}{37}{3}{9}
      \ggmoy{37}{43}{9}
      \ggmoy{43}{49}{9}
      \eemoy{49}{55}{9}
      \ccorreldeuxc{37}{-9}{37}{-4}
      \ccorreldeuxc{37}{-2}{37}{2}
      \ccorreldeuxc{37}{4}{37}{9}
      \ccorreldeuxc{43}{-9}{43}{-4}
      \ccorreldeuxc{43}{-2}{43}{2}
      \ccorreldeuxc{43}{4}{43}{9}
      \ccorreldeuxc{49}{-9}{49}{-4}
      \ccorreldeuxc{49}{-2}{49}{2}
      \ccorreldeuxc{49}{4}{49}{9}
      \pparticule{37}{9}
      \pparticule{43}{9}
      \pparticule{49}{9}
      \gggmoy{29}{37}{-3}{-9}
      \ggmoy{37}{43}{-9}
      \ggmoy{43}{49}{-9}
      \eemoy{49}{55}{-9}
      \pparticule{37}{-9}
      \pparticule{43}{-9}
      \pparticule{49}{-9}
      \nnonlineaire{29}{-3}
      \nnonlineaire{29}{3}
      \rput(58,9){$E_0$}
      \rput(58,3){$E_0$}
      \rput(58,-3){$E_0^*$}
      \rput(58,-9){$E_0^*$}
   \end{ddddiag}
\end{equation}
where the squares denote the second harmonic processes. This diagram confirms that a transport equation for the SHG beam
is still valid but with a source term given by the product of two specific intensities at $\omega$. In terms of an
equation, this source term is indeed given by
\begin{multline}
   S(\bm{r},\bm{\uprho},2\omega)
      =\int\bra G(\bm{r}-\bm{r}',2\omega)\ket\bra G^*(\bm{\uprho}-\bm{\uprho}',2\omega)\ket
\\\times
         \GammaSHG(\bm{r}',\bm{r}'',\bm{r}''',\bm{\uprho}',\bm{\uprho}'',\bm{\uprho}''',2\omega)
\\\times
         \bra E(\bm{r}'',\omega)E^*(\bm{\uprho}'',2\omega)\ket
         \bra E(\bm{r}''',\omega)E^*(\bm{\uprho}''',2\omega)\ket
\\\times
         \ud\bm{r}'\ud\bm{r}''\ud\bm{r}'''\ud\bm{\uprho}'\ud\bm{\uprho}''\ud\bm{\uprho}'''
\end{multline}
where $\GammaSHG$ is the SHG vertex given by
\begin{multline}\label{gamma_shg}
   \GammaSHG(\bm{r},\bm{r}',\bm{r}'',\bm{\uprho},\bm{\uprho}',\bm{\uprho}'',\omega)
      =\bra\chi^{(2)}(\bm{r},\omega)\chi^{(2)*}(\bm{\uprho},\omega)\ket_c
\\\times
      \delta(\bm{r}-\bm{r}')\delta(\bm{r}-\bm{r}'')\delta(\bm{\uprho}-\bm{\uprho}')\delta(\bm{\uprho}-\bm{\uprho}'').
\end{multline}
In this last expression, we keep the complex conjugate notation for the second-order susceptibility $\chi^{(2)}$ although it
is a real quantity in order to remind that it corresponds to the complex conjugate field. In the Fourier domain and
making use of the statistical homogeneity and isotropy again, we get
\begin{multline}
   \widetilde{\GammaSHG}(\bm{k},\bm{k}',\bm{k}'',\bm{\upkappa},\bm{\upkappa}',\bm{\upkappa}'',\omega)
      =8\pi^3\delta(\bm{k}-\bm{k}'-\bm{k}''-\bm{\upkappa}+\bm{\upkappa}'+\bm{\upkappa}'')
\\\times
         \GammaSHG(\bm{k},\bm{k}',\bm{k}'',\bm{\upkappa},\bm{\upkappa}',\bm{\upkappa}'',\omega).
\end{multline}
From this, we can define a SHG phase function given by
\begin{multline}
   \alpha\pSHG(\bm{u},\bm{u}',\bm{u}'',\omega)=\frac{1}{125\pi^5}
      \widetilde{\GammaSHG}(k_r(2\omega)\bm{u},k_r(\omega)\bm{u}',
\\
         k_r(\omega)\bm{u}'',k_r(2\omega)\bm{u},k_r(\omega)\bm{u}',k_r(\omega)\bm{u}'',\omega).
\end{multline}
$\alpha$ is a coefficient that takes into account all constants involved in the second harmonic generation and is such that
the second harmonic phase function is normalized as
\begin{equation}
   \int \pSHG(\bm{u},\bm{u}',\bm{u}'',\omega)\ud\bm{u}'\ud\bm{u}''=1.
\end{equation}
Plugging the expression of the correlation function $C_{\chi}$ leads to
\begin{equation}
   \pSHG(\bm{u},\bm{u}',\bm{u}'',\omega)\propto\exp\left[-\frac{q^2\ell^2}{2}\right]
\end{equation}
where $\bm{q}=k_r(2\omega)\bm{u}-k_r(\omega)\bm{u}'-k_r(\omega)\bm{u}''$ is the SHG scattering vector. We finally end up
with the non-linear RTE given by
\begin{multline}
   \left[\frac{1}{v_E(2\omega)}\frac{\partial}{\partial t}
      +\bm{u}\cdot\bm{\nabla}_{\bm{r}}+\frac{1}{\ell_e(2\omega)}\right]I(\bm{r},\bm{u},t,2\omega)
\\
      =\frac{1}{\ell_s(2\omega)}\int p(\bm{u}\cdot\bm{u}',2\omega)I(\bm{r},\bm{u}',t,2\omega)\ud\bm{u}'
\\
      +\alpha\iint \pSHG(\bm{u},\bm{u}',\bm{u}'',\omega)I(\bm{r},\bm{u}',t,\omega)I(\bm{r},\bm{u}'',t,\omega)\ud\bm{u}'\ud\bm{u}''
\end{multline}
which is Eq.~(\ref{non_linear_rte}).

\end{document}